\documentclass[preprint,nofootinbib,aps,superscriptaddress]{revtex4-1}
\usepackage{graphicx}% Include figure files
\usepackage{dcolumn}% Align table columns on decimal point
\usepackage{bm}% bold math
\usepackage{latexsym}
\usepackage{amsfonts}
\usepackage{amssymb}
\usepackage{amsmath}
\usepackage[usenames]{color}
\begin{document}
\preprint{KUNS-2627}
\preprint{RESCEU-21/16}

\title{Primordial fluctuations from inflation in dRGT bimetric theory of gravity}
\author{Yuki Sakakihara}
\affiliation{Department of Physics, Kyoto University, Kyoto 606-8502, Japan}
\affiliation{Research Center for the Early Universe (RESCEU), Graduate School of Science,
The University of Tokyo, Tokyo 113-0033, Japan}
\author{Takahiro Tanaka}
\affiliation{Department of Physics, Kyoto University, Kyoto 606-8502, Japan}
\affiliation{Yukawa Institute for Theoretical Physics, Kyoto University, Kyoto 606-8502, Japan}

%\date{\today}

\begin{abstract}
We investigate primordial gravitational waves and curvature perturbations in de~Rham-Gabadadze-Tolley (dRGT) bimetric gravity. We evaluate the power-spectra in the leading order in slow roll. Taking into account the decay of massive graviton, we find that the action up to the second order reduces to the Einstein theory with a non-minimally coupled scalar field, which is simplified to a minimally coupled model by conformal transformation. We also find that the tensor to scalar ratio for large field inflation with power law potential is larger than the general relativity counterpart for any choice of parameters in dRGT bimetric gravity.
In addition, we confirm that the usual consistency relation holds and we have a steeper spectrum for gravitational waves.
%, while the spectral index of scalar power-spectrum can be either larger or smaller than the counterpart, depending on e-folding number, power law index of the potential and the parameters in the bimetric interaction.
\end{abstract}

%%\pacs{}% PACS, the Physics and Astronomy
%                             % Classification Scheme.
%%\keywords{Suggested keywords}%Use showkeys class option if keyword
%                              %display desired

\maketitle
%%\tableofcontents

\section{Introduction} 

%We have calculated the spectrum of the primordial gravitational waves, however, it is not easy to compare the result with observations since we have not observe the primordial gravitational waves yet due to the smallness of the amplitude. Furthermore,
%\begin{eqnarray}
% r=\frac{\mathcal{P}_T}{\mathcal{P}_{\mathcal{R}}} \ ,
%\end{eqnarray}
%\begin{eqnarray}
% n_s-1=\frac{d \log \mathcal{P}_\mathcal{R}}{d \log k} \ ,
%\end{eqnarray}
%, which is the ratio of the spectrum of the primordial gravitational waves to the spectrum of the primordial curvature perturbations 
%, which is the scale dependence of the curvature perturbations 

The inflation scenario predicts primordial gravitational waves and curvature perturbations stemming from the quantum fluctuations of the inflaton. The tensor to scalar ratio and the spectral index of the curvature perturbation are parameters constrained by current and future observations of the cosmic microwave background. The power law shape of the inflaton potential written as $V[\phi]\propto \phi^n$ with a positive integer $n$ gives a typical slow roll inflation model. According to the latest results of Planck observations~\cite{Ade:2015xua, Ade:2015lrj}, the quartic potential $\propto\phi^4$ is almost excluded and the quadratic potential $\propto\phi^2$ is also out of the $2\sigma$ confidence region on the $n_s$-$r$ plane. 

%\textcolor{green}{As a clew for understanding unknown components of the universe, such as dark energy and dark matter, infrared modification of gravity has been discussed.}
%\footnote{What I wrote here means just a motivation of modified gravity theories in general. I thought we should explain why we consider such a modification of theory. However, the explanation may be unnecessarily as you said.} 
Possibility that gravitons are massive is one of natural modifications of gravity theory and is worthwhile to be considered. Recent progress in massive gravity is driven by the construction of a consistent model~\cite{deRham:2010ik,deRham:2010kj,Hassan:2011hr,Hassan:2011zd,Hassan:2011ea}, which is free from the Boulware-Deser ghost~\cite{Boulware:1973my}. The model only with a massive graviton, however, has no stable homogeneous isotropic cosmological solution~\cite{DeFelice:2013awa} and general covariance is manifestly violated. Extension to bimetric theory of gravity with an additional metric recovers general covariance and, moreover, we can easily have stable homogeneous isotropic cosmological solutions~\cite{Volkov:2011an, vonStrauss:2011mq, Comelli:2012db, Berg:2012kn, Sakakihara:2012iq, Volkov:2013roa, DeFelice:2013nba, Konnig:2014xva, Lagos:2014lca, Cusin:2014psa}. At least, at low energies compared with the energy scale of the interaction terms between two metrics, there is a cosmological background solution which is free from ghost and gradient instabilities~\cite{DeFelice:2013nba,DeFelice:2014nja}.
%(Black holes~\cite{} and the realization of general relativity within the solar system~\cite{} have also been discussed in bimetric gravity.) 
Since bimetric theory has not only a massive graviton, which decays rapidly during inflation, but also a massless graviton, which survives in contrast, we can have a spectrum of primordial gravitational waves similar to the one predicted by general relativity~\cite{Sakakihara:2015naa}.~(See also~\cite{Cusin:2015pya}.)

The aim of this paper is to reveal whether bimetric theory can be consistent with observations or not and, if the answer is ``yes'', how the theory is constrained by observations. We adopt the simplest setup of inflation where a scalar field is coupled to the physical metric as the inflaton. In this paper, we extend the previous results on primordial gravitational waves in the following sense: (i) to general model parameters of dRGT bimetric theory at the leading order in slow roll, and (ii) to primordial curvature perturbations. The latter extension is essential in order to compare the results with the current constraints on inflation models obtained from cosmic microwave background observations. In particular, we discuss how the constraints on the inflaton potential for the single-field slow roll inflation are modified. We can easily anticipate that, for some models, the already tight constraints on the power law potentials might be relaxed due to the effects of modified gravity. Contrary to the naive expectation, we will find that the constraint on the tensor to scalar ratio becomes even tighter in any choice of the bimetric model parameters, except when the graviton's mass is very close to the Higuchi bound~\cite{Higuchi:1986py, Sakakihara:2012iq, DeFelice:2014nja}. This exceptional case is not discussed in detail here.

We organize this paper as follows. In Sec.~2, we show the general action of dRGT bimetric model of gravity with an inflaton. In Sec.~3, we derive background equations in this model and define slow roll parameters characterizing the inflaton potential. In Sec.~4, we argue that the perturbed action can be simplified in the slow roll approximation to evaluate the late time spectra of cosmological perturbations owing to the decay of massive gravitons. In Sec.~5, we calculate the spectra of the primordial tensor and curvature perturbations with the aid of the reduced action to the leading order in slow roll. In Sec.~6, as a concrete example, we evaluate how the tensor to scalar ratio and the spectral index of the curvature perturbation are modified from the general relativity counterparts for the power law potentials. Section~7 is devoted to the summary of our results.

\section{Action}

The bimetric action with a scalar field $\phi$ coupled to the physical metric $g_{\mu\nu}$,
which is considered as the inflaton, is~\footnote{We mention that there is another description including the metric description of bimetric theory, which is referred to as the vierbein formulation~\cite{Hinterbichler:2012cn, Hassan:2012wt, Deffayet:2012zc, Tamanini:2013xia, Hinterbichler:2015yaa, deRham:2015rxa, deRham:2015cha}.}
%We treat the general bimetric action without fixing the bimetric parameters.
%, though we have examined the minimal bimetric action so far. 
\begin{eqnarray} 
   S&=&\frac{M_{g}^{2}}{2}\int d^{4}x \sqrt{-g}R[g_{\mu\nu}]
    +\int d^{4}x \sqrt{-g}\Bigl(-\frac{1}{2}g^{\mu\nu}\partial_{\mu}\phi\partial_{\nu}\phi-V[\phi]\Bigr)\nonumber \\
    &&+\frac{M_{f}^{2}}{2}\int d^{4}x \sqrt{-f}R[f_{\mu\nu}]
    -m^{2}M_{g}^{2}\int d^{4}x \sqrt{-g} \,\sum_{k=0}^4 c_k F_k[Y^\mu_\nu]\ ,
\label{action_w_scalar}
\end{eqnarray}
where $m$ is the coupling constant between the metrics. 
$\{c_k\}$ are bimetric model parameters, which are basically free parameters.
$M_g$ and $M_f$ are the gravitational coupling energy scales of $g$-metric and $f$-metric, respectively.
$R[g_{\mu\nu}]$ is the Ricci scalar constructed from $g$-metric and $R[f_{\mu\nu}]$ is defined in a similar manner. $V[\phi]$ is the potential of the scalar field.
$Y^\mu_\nu$ is defined by 
$Y^\mu_\alpha Y^\alpha_\nu
=g^{\mu\alpha}f_{\alpha\nu}$, and 
\begin{equation}
 F_{k}[X^{\mu}_{\nu}]
 =\frac{1}{k!}\sum_{\sigma\in S_{k}}\mathrm{sgn}(\sigma)
 X_{\mu_{1}}^{\mu_{\sigma(1)}}X_{\mu_{2}}^{\mu_{\sigma(2)}}\cdots
 X_{\mu_{k}}^{\mu_{\sigma(k)}} \ ,
\label{Fn_def}
\end{equation}
where $S_k$ is the permutation group of degree $k$ and sgn$(\sigma)$ is
$+1$ for even number permutations while $-1$ for odd number permutations. 
More explicitly, the functions $F_k$ are written as
\begin{eqnarray}
 &&F_0[X^{\mu}_{\nu}]=1 \ , \quad F_1[X^{\mu}_{\nu}]=[X] \ , \quad F_2[X^{\mu}_{\nu}]=\frac{1}{2}([X]^2-[X^2]), \nonumber\\
 &&F_3[X^{\mu}_{\nu}]=\frac{1}{6}([X]^3-3[X][X^2]+2[X^3]) \ , \quad F_4[X^\mu_\nu]=\det(X) \ ,
\end{eqnarray}
where $[X^k]$ on the right hand sides means the trace of $X^k$.
%It is needed that matter coupled to $g$ 

We called $g$-metric the physical metric since it is directly coupled to the ordinary matter fields. The scalar field is also assumed to be coupled to $g$-metric so as to decay to the ordinary matter fields during reheating since the decay through gravitational couplings will be inefficient if the scalar field is coupled only to $f$-metric. Of course, the scalar field can be coupled to both of $g$-metric and $f$-metric~\cite{Akrami:2013ffa, Yamashita:2014fga, deRham:2014naa}. One possibility is the potential couplings. Another possibility is the kinetic coupling through the effective metric constructed from $g$-metric and $f$-metric. Otherwise, Boulware-Deser ghost reappears. These cases are not considered and we discuss the simplest case that the scalar field is coupled only to the physical metric.

\section{Inflationary spacetime in bimetric theory}\label{INF_SP_BI}

Inflationary background spacetime in bimetric theory is described by 
\begin{align}
 ds^2\equiv g_{\mu\nu}dx^\mu dx^\nu
=-N^2 dt^2+a^2 (dx^2+dy^2+dz^2) \ ,
\end{align}
\begin{align}
 ds'{}^{2}\equiv f_{\mu\nu}dx^\mu dx^\nu=- M^2 dt^2+ b^2 (dx^2+dy^2+dz^2) \ .
\end{align}
%where we decomposed the scale factor of $f$-metric $b$ as $b=\xi a$ by defining $\xi=b/a$.
After substitution of them into the action~\eqref{action_w_scalar}, we
find that the action does not contain time derivatives of $N$ and 
$M$. 
From the variations with respect to them, we obtain two constraints:~\cite{DeFelice:2014nja}
\begin{align}
 H^2\equiv\Bigl(\frac{\dot{a}}{a}\Bigr)^2=\frac{1}{3M_g^2}\biggl[U-\frac{\xi}{4}U'+\rho_{\phi}\biggr] \ ,
\label{const_H}
\end{align}
\begin{align}
 H_f^2\equiv\Bigl(\frac{N \dot{b}}{M b}\Bigr)^2=\frac{U'}{12\kappa\xi^3 M_g^2} \ ,
\label{const_Hf}
\end{align}
and, from the variations with respect to dynamical variables, three equations of motion:
\begin{align}
 \dot{H}=\frac{(M/N-\xi)J}{6M_g^2}-\frac{\rho_\phi+p_{\phi}}{2M_g^2} \ ,
\label{eom_phys}
\end{align}
%\begin{align}
% \dot{H}=\frac{\xi(\tilde{c}-1)}{6M_g^2}J-\frac{\rho_\phi+p_{\phi}}{2M_g^2} \ ,
%\label{eom_phys}
%\end{align}
%\begin{align}
% \dot{H}=\frac{\xi(\tilde{c}-1)}{6M_g^2}J-\frac{\dot{\phi}^2}{2M_g^2} \ ,
%\label{eom_phys}
%\end{align}
\begin{align}
 \dot{H_f}=-\frac{(M/N-\xi)J}{6\kappa\xi^3M_g^2} \ ,
\label{eom_fid}
\end{align}
%\begin{align}
% \dot{H_f}=-\frac{\xi(\tilde{c}-1)}{6\kappa\xi^3M_g^2}J \ ,
%\label{eom_fid}
%\end{align}
\begin{align}
 \ddot{\phi}+3H\dot{\phi}+V_{\phi}=0 \ ,
\label{eom_phi}
\end{align}
where $\rho_\phi$ is the matter energy density, $p_\phi$ is the pressure of matter and
\begin{equation}
 \dot{~}\equiv\frac{d}{Ndt} \ .
\end{equation}
We introduced also the ratio of two gravitational couplings $\kappa\equiv M_f^2/M_g^2$ and the ratio of two scale factors $\xi\equiv b/a$. $U$ and $J$ are functions of $\xi$, defined by
\begin{align}
 U(\xi)=m^{2}M_{g}^{2}(c_0+4c_1\xi+6c_2\xi^2+4c_3\xi^3+c_4\xi^4) \ ,
\end{align}
and
\begin{align}
 J(\xi)=\biggl[U-\frac{\xi}{4}U'\biggr]'=3m^2M_g^2(c_1+2c_2\xi+c_3\xi^2) \ ,
\label{def_J}
\end{align}
respectively.
In order for the constraints \eqref{const_H} and \eqref{const_Hf} to be satisfied after the time evolution using Eqs.~\eqref{eom_phys}, \eqref{eom_fid} and \eqref{eom_phi}, an additional constraint,
\begin{align}
 J(H-H_f\xi)=0 \ ,
\end{align}
is required. Although there is still a room for debate,
the branch with $J=0$,
which lacks scalar (=helicity-0) and vector (=helicity-1) mode perturbations 
at the level of linear perturbation, is likely to be pathological in general~\cite{Gumrukcuoglu:2011zh,Comelli:2012db,DeFelice:2012mx,Gumrukcuoglu:2012aa,Tasinato:2012ze}.
Here in this paper, we concentrate on the background that satisfies 
\begin{align}
 H=H_f \xi \ ,
\label{health_branc}
\end{align}
which we call the healthy branch.
With the additional constraint, we can rewrite the two constraints~\eqref{const_H} and \eqref{const_Hf} as 
\begin{align}
 H^2=\frac{U+\rho_\phi}{3(1+\kappa\xi^2)M_g^2}=\frac{U'}{12\kappa \xi M_g^2}\ .
\label{Friedmann}
\end{align}
The latter equality can be rearranged as 
\begin{align}
 F=\rho_\phi=\frac{1}{2}\dot{\phi}^2+V[\phi] \ ,
\label{det_xi}
\end{align}
where
\begin{align}
 F(\xi)=-U+\frac{(1+\kappa\xi^2)}{4\kappa\xi}U' \ .
\label{xi_det}
\end{align}
We can read off the effective Planck constant from the Friedmann equation~\eqref{Friedmann}
as
\begin{align}
 M_{\rm eff}^2=\frac{(1+\kappa\xi^2)}{(1+U/\rho_\phi)}M_g^2  \ .
\end{align}
 The ratio of the scale factors $\xi$ is determined by $\rho_\phi$ through Eq.~\eqref{det_xi}. 

We introduce $\tilde{c}$ as $\tilde{c}=M/\xi N$, which describes the difference between two light cones. We can show the relation between $\tilde{c}$ and $\xi$,
\begin{align}
 \tilde{c}-1=\frac{\dot{\xi}}{H\xi} \ ,
\label{tildec_xi}
\end{align}
from the definition of $\xi$ and the constraint \eqref{health_branc}.
The time derivative of \eqref{health_branc} followed by Eqs.~\eqref{eom_phys} and \eqref{eom_fid} gives us
\begin{align}
 \tilde{c}-1=\frac{\rho_\phi+p_\phi}{M_g^2 W}=\frac{\dot{\phi^2}}{M_g^2 W} \ ,
\label{tildc_phi}
\end{align}
where 
\begin{align}
 W(\xi)=\frac{(1+\kappa\xi^2)}{3\kappa\xi M_g^2}J-2H^2 \ .
\label{def_WJ}
\end{align}
We note that the absence of the Higuchi ghost requires $W>0$~\cite{DeFelice:2014nja} and, consequently, $\tilde{c}$ is larger than unity and $\dot{\xi}$ is positive. In de Sitter case, where energy density of matter is constant and $\rho_\phi+p_\phi=0$ is satisfied, we have a constant $\xi$ and $\tilde{c}$ is just unity. Furthermore, we need to assume the coupling constant, $m$, is larger than the Hubble scale, i.e., $m\gtrsim H$, after inflation in order to avoid gradient instability~\cite{DeFelice:2014nja}.
%We introduce $\tilde{c}$ as $\tilde{c}=M/\xi N$, which describes the difference between two light corns. Combining Eqs. (\ref{eom_phys}), (\ref{eom_fid}) and (\ref{health_branc}), we find
%\begin{align}
% \tilde{c}-1=\frac{\rho_\phi+p_\phi}{M_g^2 W}=\frac{\dot{\phi^2}}{M_g^2 W} \ ,
%\label{tildc_phi}
%\end{align}
%where 
%\begin{align}
% W(\xi)=\frac{(1+\kappa\xi^2)}{3\kappa\xi M_g^2}J-2H^2 \ .
%\label{def_WJ}
%\end{align}
%We note that the absence of the Higuchi ghost requires $W>0$ \cite{DeFelice:2014nja} and, consequently, $\tilde{c}$ is larger than unity. 
%$\tilde{c}$ and the time derivative of $\xi$ are related as
%\begin{align}
% \tilde{c}-1=\frac{\dot{\xi}}{H\xi} \ ,
%\label{tildec_xi}
%\end{align}
%which is derived from the constraint (\ref{health_branc}).

%We also obtain an usual equation of motion for the scalar field

A cosmological solution can be realized with $\xi$ chosen as a root of Eq.~\eqref{xi_det} depending on the energy density of matter. The number of the roots are determined by the parameters $\{c_k\}$ and $\kappa$. As we will see later, the generalized Higuchi bound is reduced to the condition $F'<0$, which is easily realized. We also have the freedom to set the root of $F=0$, $\xi_0$, to unity with the following transformation: $\xi\rightarrow\xi/\xi_0$, $c_k\rightarrow c_k \xi_0^n$, and $\kappa\rightarrow \kappa \xi_0^2$. Therefore, even if there are several different roots with the same model parameters, they can be absorbed by different choices of the model parameters.
%though the root seems unique at least under the assumptions that the Higuchi bound and $m\gg H$ are satisfied

%For instance, we have at least a root satisfying the Higuchi bound when $c_1>0$ and $c_3>0$, which imply $F\rightarrow \infty$ as $x\rightarrow 0$ and $F\rightarrow -\infty$ as $x\rightarrow \infty$, respectively. 

We use the slow roll approximation in the following discussion.
We introduce slow roll parameters by 
\begin{align}
 \epsilon=\frac{M_{\rm eff}^2}{2}\Bigl(\frac{V_\phi}{V}\Bigr)^2 \ , \qquad
 \eta=M_{\rm eff}^2\frac{V_{\phi\phi}}{V} \ .
\label{slow_param}
\end{align}
%\begin{align}
% H^2=\frac{V+U}{3(1+\kappa\xi^2)M_g^2}=\frac{V}{3M_{\rm eff}^2} \ ,
%\label{Fried_approx}
%\end{align}
Under the slow roll conditions, $\epsilon$, $\left|\eta\right| \ll 1$, 
the background equations reduce in the leading order to
\begin{equation}
 H^2=\frac{V}{3M_{\rm eff}^2} \ , \qquad 
 3H\dot{\phi}+V_\phi=0 \ , \qquad 
 F=V\ ,
\label{slow_back}
\end{equation}
where
\begin{equation}
  M_{\rm eff}^2=\frac{1+\kappa\xi^2}{1+U/V}M_g^2 \ .
\label{eff_def}
\end{equation}
From Eqs.~\eqref{slow_param} and \eqref{slow_back}, the time derivative of the scalar field is evaluated as 
\begin{equation}
 \dot{\phi}=-\sqrt{2\epsilon}M_{\rm eff} H \ .
\label{phi_dot}
\end{equation}
Therefore, from Eq.~\eqref{tildc_phi}, we obtain
\begin{align}
 \tilde{c}-1=\frac{2\epsilon M_{\rm eff}^2H^2}{M_g^2 W}=\frac{2\epsilon V}{3M_g^2 W}\,,
\label{tildec_W}
\end{align}
i.e., the deviation of $\tilde{c}$ from unity is suppressed as 
$\mathcal{O}(\epsilon)$ under the slow roll approximation, 
unless $W$ is tuned to be extraordinary small compared with 
$V/M_g^2\approx H^2$. 
Also, $\dot{\xi}/\xi$ has the same suppression 
according to Eq.~\eqref{tildec_xi}.

Here we briefly mention the exceptional case. If we choose such parameters exhibiting $W\ll V/M_g^2$, namely, the Higuchi bound is saturated, the analysis we perform is not valid any more even in the background level. This is because the other light cone largely deviates from the physical one and the relation between the spatial scales $\xi$ rapidly changes in spite of slow roll of the inflaton, which means the inflationary spacetime is not well-approximated by de Sitter spacetime and the slow roll limit cannot be taken. That behavior can be seen in the right hand side of \eqref{tildec_W} enhanced due to the smallness of $W$. We exclude this finely tuned situation in the following discussion. 

%When we calculate the spectra under slow roll conditions, the apparent dependence on slow roll parameters included in $\tilde{c}$ is the next leading effects.
%, while the dependence of $\xi$ on the scalar field can contribute as a leading effect. 
%Since the correction to the spectra is sub-

\section{Cosmological perturbations}\label{Cosmo_Pert}

In the slow roll limit, $\epsilon$, $\eta \to 0$, tensor (=helicity-2) 
modes are decomposed into massless modes and massive modes. The massless modes obey usual equations of motion and the massive modes follow the equations of motion with the effective mass~\cite{DeFelice:2014nja},
\begin{equation} 
 m_{\rm eff}^2=\frac{(1+\kappa\xi^2)}{3\kappa \xi M_g^2}J \ .
\label{eff_mass}
\end{equation}
On the other hand, there is no massless dynamical gravitational 
degree of freedom in the scalar modes, 
but the massive graviton has a scalar mode. 
On de Sitter background, which is realized in the slow roll limit, 
three different helicity modes of massive graviton take the same effective mass~\eqref{eff_mass}. It is known that 
a factor $W=m_{\rm eff}^2 -2H^2$ appears in front of the kinetic term 
of the scalar massive graviton, and hence a ghost instability 
takes place if $m_{\rm eff}^2 \leq 2H^2$~\cite{Higuchi:1986py,Fasiello:2012rw}, which is referred as the Higuchi ghost. 
(More precisely, the strong coupling occurs in the perturbative expansion when $m_{\rm eff}^2=2H^2$.)
Therefore, we assume that the model parameters are chosen so that 
$m_{\rm eff}^2 > 2H^2$ is satisfied. 
In the slow roll limit,
perturbations in the matter sector are decoupled from the
gravitational sector, obey the equation of motion
of a massless field and have a flat spectrum as in general
relativity. As a result, in the present setup, both tensor perturbations and
scalar perturbations have massive modes and massless modes. The
massive modes are heavy enough to decay during inflation just like usual
matter without pressure. Therefore they are irrelevant for the 
generation of seeds of structure formation during inflation.
Only the massless modes contribute to 
the perturbation spectra observed after the inflation.

\begin{figure}[!h]
\centering \includegraphics[height=3.0in]{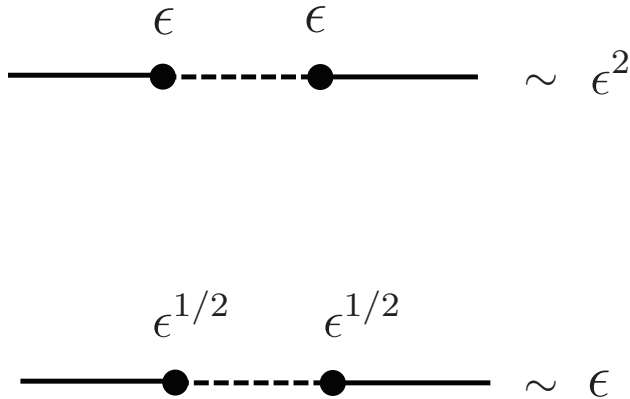}
 \caption{The leading corrections to the tensor and scalar spectra including massive propagators are shown. The upper diagram is the leading correction to the tensor spectrum. The solid line means the propagation of massless graviton and the dashed line means the propagation of massive graviton there. The vertex is $\mathcal{O}(\epsilon)$ because of the time derivative of $\xi$ and the deviation of $\tilde{c}$ from unity. The lower diagram is the leading correction to the scalar spectrum. The solid line  means the propagation of massless scalar modes and the dashed line again means the propagation of massive graviton there. The vertex is $\mathcal{O}(\epsilon^{1/2})$, which leading contribution comes from $\dot{\phi}$.}
\label{Fig1}
\end{figure}

Under the slow roll approximation, the mixing between 
the massless tensor modes and the massive tensor modes 
appears in $\mathcal{O}(\epsilon)$. 
This is because the time derivative of $\xi$ 
and the deviation of $\tilde{c}$ from unity are involved in 
the mixing terms, and they are $\mathcal{O}(\epsilon)$. 
Therefore, the leading order correction to the amplitude of the tensor
spectrum comes from the diagram schematically shown in Fig.~\ref{Fig1}, 
where a solid line represents the massless graviton propagator and 
a dashed line represents the massive graviton propagator. 
The external lines should end with a massless graviton since 
the amplitude decays otherwise. 
Since this diagram contains two slow roll suppressed vertices, 
the contribution becomes $\mathcal{O}(\epsilon^2)$. 
As we concentrate on the leading order effects, we neglect 
such higher order corrections. 

The massive scalar modes
and the scalar field modes are also coupled when we expand the perturbations
with respect to the slow roll parameters. 
In contrast to the tensor modes, the mixing vertex in the scalar modes 
appears in $\mathcal{O}(\epsilon^{1/2})$
because the leading order terms are slow roll suppressed only by $\dot{\phi}$. 
Hence, the correction to the amplitude of the scalar spectrum can appear in $\mathcal{O}(\epsilon)$, which is also a sub-leading effect in slow roll. 
The above discussion allows us to ignore slow roll suppressed terms in the perturbed action. Then, the modifications on the spectra coming from the massive modes, which decay rapidly, can be simply neglected, and therefore we are allowed to assume 
\begin{equation}
 f_{\mu\nu}=\xi^2 g_{\mu\nu} \ ,
\label{reduc_xi}
\end{equation}
as long as we are concerned with the leading order effects in the amplitude of the spectra and
the spectral index, where $\xi$ is determined by the value of $\phi$ through the last equation of \eqref{slow_back}. The difference of $\xi$ from the background value is estimated as $\xi[\phi]= \xi[\phi_0]+(d\xi/d\phi)|_{\phi_0} \delta \phi$, where $\phi_0$ and $\delta \phi$ are the background value and the perturbation of $\phi$, respectively. $(d\xi/d\phi)|_{\phi_0}$ is slow roll suppressed as $\epsilon^{1/2}$, since $\dot{\phi_0}$ is $\mathcal{O}(\epsilon^{1/2})$ and $\dot{\xi}/\xi$  is $\mathcal{O}(\epsilon)$,
%respectively as you can see from Eq.~\eqref{tildec_xi}, \eqref{phi_dot} and \eqref{tildec_W},
which we saw in Sec.~\ref{INF_SP_BI}. Therefore, the perturbation of $\xi$ arising from $\delta \phi$ is neglected.
%\eqref{det_xi}
%Though $\xi$ can contribute as a massive mode even after we assume the relation \eqref{reduc_xi}, taking into account the decay of the massive modes, we can treat $\xi$ as a non-dynamical degree of freedom.
%Here we also implicitly assumed that the 
We can insist that the manifestly slow roll suppressed corrections are not
enhanced since we have discarded the situation that the Higuchi bound is saturated.
%(The corrections can be enhanced due to the rapid change of
%the ratio of two scale factors $\xi$ despite the slow motion of the inflaton, 
%which occurs when the Higuchi bound is saturated. We briefly mention that case later.)

We then obtain an action similar to the Einstein theory with a non-minimally coupled scalar field. 
\begin{align} 
   S=&\frac{M_{g}^{2}}{2}\int d^{4}x \sqrt{-g}R[g_{\mu\nu}]
    +\frac{M_{g}^{2}}{2}\int d^{4}x \sqrt{-g} \,\kappa\xi^2(R[g_{\mu\nu}]+6g^{\mu\nu}\partial_\mu \log \xi \,\partial_\nu \log \xi )\nonumber \\
    &+\int d^{4}x \sqrt{-g}\Bigl(-\frac{1}{2}g^{\mu\nu}\partial_{\mu}\phi\,\partial_{\nu}\phi-V[\phi]\Bigr)
     -\int d^{4}x \sqrt{-g} U(\xi) \ .
\label{Mod_1}
\end{align}
%Though $\xi$ is a function of homogeneous background part of $\phi$, it can be considered just as a function of $\phi$. It is justified by the following discussion.
%$(d\xi/d \phi)|_{\phi_0}$ is $\mathcal{O}(\epsilon^{1/2})$. Therefore, $\xi[\phi]=\xi[\phi_0]$ holds in the leading order in slow roll.
%$\xi$ is denoted by a function of $\phi$ instead of the background part of $\phi$. 
%It is justified since the perturbations arising from $\xi[\phi]$ are sub-leading in slow roll, which is easily understood through  
We can see that $(\partial \log \xi)^2$ is slow roll suppressed as $\epsilon$ compared to $(\partial \phi)^2$ as follows.
%More explicitly, 
\begin{align}
 \frac{d \log \xi}{d \phi}=\frac{1}{\xi}\frac{d\xi}{d\phi}
=\frac{V_\phi}{\xi F'}
=-\frac{V_\phi}{3M_g^2W}=-\frac{\sqrt{2\epsilon}}{M_{\rm eff}}\frac{V}{3M_g^2W} \ ,
\label{logO_phi}
\end{align}
where we have used a background equation in \eqref{slow_back}, the relation between $F'$ and $W$,
\begin{equation}
 F'=-\frac{3M_g^2 W}{\xi} \ , 
\label{rel_FW}
\end{equation}
found by combining Eqs.~\eqref{def_J}, \eqref{Friedmann}, \eqref{xi_det} and \eqref{def_WJ}, and the definition of $\epsilon$. %This means that $\xi$ moves slower than $\phi$.
The suppression is effective and $(\partial \log \xi)^2$ can be neglected in the action~\eqref{Mod_1},
%$(\partial \log \xi)^2$ in \eqref{Mod_1} is slow roll suppressed as $\epsilon$ compared to $(\partial \phi)^2$ and can be neglected. It works 
as long as $W$ is not exceptionally small compared to $V/M_g^2$, which we already required at the end of Sec.~\ref{INF_SP_BI}.
%is satisfied under our assumption that $W$ is not so small compared to $H^2$ as we mentioned 
%We briefly mention the sign of $F'$. 
%We can find that 
%$F'$ is related to $W$ as
From Eq.~\eqref{rel_FW}, the positivity of $W$ required for the absence of the Higuchi ghost is equivalent to 
\begin{equation}
 F'<0 \ , 
\label{Higuchi_condition}
\end{equation}
which we use in Sec.~\ref{Power-Law}. We also see $d\xi/d\phi<0$ from \eqref{logO_phi}.
%We also find that the rapid change of $\xi$ occurs even in slow roll when $W/H^2\ll 1$, which means the Higuchi bound is saturated, referring Eq.~\eqref{logO_phi}. In the exceptional case, the slow roll limit of inflationary spacetime cannot be taken any more and another analysis is needed at the background level. We do not discuss that case here and think it as a succeeding work.
The action reduces to
\begin{align} 
   S=\frac{M_{g}^{2}}{2}\int d^{4}x \sqrt{-g}(1+\kappa \xi[\phi]^2)R[g_{\mu\nu}]
    +\int d^{4}x \sqrt{-g}\Bigl(-\frac{1}{2}g^{\mu\nu}\partial_{\mu}\phi\,\partial_{\nu}\phi-(V[\phi]+U(\xi[\phi]))\Bigr) \ .
\label{Mod_2}
\end{align}
This is nothing but the Einstein theory with a non-minimal coupling scalar field. The equations of motion \eqref{slow_back} are obtained even if we forget that $\xi$ is a function of $\phi$ and consider $\xi$ as an independent variable under the slow roll approximation.
% In the latter case, combined with the Friedmann equation, the variation of the action with respect to $\xi$ turns out to be equivalent to \eqref{det_xi}. 
%At the background level, $\xi$ can be considered either as a function of $\rho_\phi$ determined by the background relation Eq.~\eqref{det_xi} or as an independent variable. In the latter case, combined with 
%the Friedmann equation,  
%the variation of the action with respect to $\xi$ turns out to be 
%equivalent to Eq.~\eqref{det_xi}. 
%The scalar potential is effectively modified by the potential terms originating from the interaction between the metrics. 
%\begin{align} 
%   S=&\frac{M_{g}^{2}}{2}\int d^{4}x \sqrt{-g}R[g_{\mu\nu}]
%    +\frac{M_{g}^{2}}{2}\int d^{4}x \sqrt{-g} \,\kappa\xi^2(R[g_{\mu\nu}]+6g^{\mu\nu}\partial_\mu \log \xi \,\partial_\nu \log \xi )\nonumber \\
%    &+\int d^{4}x \sqrt{-g}\Bigl(-\frac{1}{2}g^{\mu\nu}\partial_{\mu}\phi\,\partial_{\nu}\phi-V[\phi]\Bigr)
%     -\int d^{4}x \sqrt{-g} U(\xi) \ .
%\end{align}

Let us make a conformal transformation 
\begin{eqnarray}
 g_{\mu\nu}=\Omega^2\bar{g}_{\mu\nu} \ , \qquad \mathrm{where} \qquad  \Omega=\Bigl(\frac{1}{1+\kappa\xi^2}\Bigr)^{1/2} \ ,
\label{conf_trans}
\end{eqnarray}
to remove the non-minimal curvature coupling term. 
In terms of the new metric $\bar{g}$ defined by the above relation, 
the action is rewritten as 
\begin{align}
 S=& \frac{M_{g}^{2}}{2}\int d^{4}x \sqrt{-\bar{g}}\bar{R}\nonumber\\
 &+\int d^{4}x \sqrt{-\bar{g}}\Bigl(-\frac{1}{2}\biggl[\Omega^2+6M_g^2\Bigl(\frac{\kappa\xi^2}{1+\kappa\xi^2}\Bigr)^2\Bigl(\frac{d \log \xi}{d \phi}\Bigr)^2\biggr]\bar{g}^{\mu\nu}\partial_{\mu}\phi\,\partial_{\nu}\phi-\Omega^4(V[\phi]+U(\xi[\phi]))\Bigr)  \ . %\nonumber\\
\end{align}
Again, we neglect $(d\log \xi/d\phi)^2$ by the same discussion as above.
We define a new scalar field to absorb the conformal factor:
\begin{equation}
 \psi\equiv\int d\phi\;\Omega  \ .
\label{phi_psi_rel}
\end{equation}
Then, we obtain the standard action with a single inflaton as 
\begin{equation}
 S=\frac{M_{g}^{2}}{2}\int d^{4}x \sqrt{-\bar{g}}\bar{R}
  +\int d^{4}x \sqrt{-\bar{g}}\Bigl(-\frac{1}{2}\bar{g}^{\mu\nu}\partial_{\mu}\psi\partial_{\nu}\psi-\Omega^4[V(\phi[\psi])+U(\xi)]\Bigr) \ .
\end{equation}
%the relation between $\phi$ and $\psi$ can be evaluated as
%\begin{equation}
% \frac{d\phi}{d\psi}=\frac{1}{\Omega} \ ,
%\label{phi_psi_rel}
%\end{equation}

To summarize, taking into account the decay of massive modes and concentrating on the leading effects in slow roll, 
%after replacing the metric and the scalar field with $\bar{g}$ and $\psi$, respectively, 
the original bimetric action reduces to the Einstein theory with a non-minimally coupled 
scalar field, which is equivalent to the minimally coupled scalar system through a conformal transformation.

\section{Primordial spectra in bimetric theory}
In this section, we derive the formulae for 
the spectral index of the scalar spectrum and 
the tensor to scalar ratio in bimetric theory.
Since the reduced action is exactly the same as the Einstein theory with a
minimally coupled scalar field, the spectra can be calculated 
following the standard way. Furthermore, metric 
perturbations after factoring out the background scale factor 
are not changed under the conformal transformation. 
Therefore, the power-spectra for the curvature perturbation and the tensor 
perturbation obtained in the reduced system are identified with those in the
original system.
These power-spectra are known to be given by 
\begin{eqnarray}
 \mathcal{P}_T=2\Bigl(\frac{\bar{H}}{\pi M_g}\Bigr)^2
\ , \qquad
 \mathcal{P}_\mathcal{R}=\frac{1}{8\bar{\epsilon}}\Bigl(\frac{\bar{H}}{\pi M_g}\Bigr)^2\ ,
\label{power_spectra}
\end{eqnarray}
where we define the Hubble expansion rate $\bar{H}$ in the reduced system, which is evaluated as
\begin{eqnarray}
 \bar{H}^2=\frac{\bar{V}}{3M_g^2} \ ,
\end{eqnarray}
and a slow roll parameter 
\begin{eqnarray}
 \bar{\epsilon}=\frac{M_{g}^2}{2}\Bigl(\frac{\bar{V}_\psi}{\bar{V}}\Bigr)^2 \ ,
\end{eqnarray}
with 
\begin{eqnarray}
 \bar{V}=\Omega^4[V+U] \ .
\label{new_pot}
\end{eqnarray}
In addition, we define another slow roll parameter as
\begin{eqnarray}
 \bar{\eta}=M_{g}^2\frac{\bar{V}_{\psi\psi}}{\bar{V}} \ .
\end{eqnarray}
The tensor to scalar ratio, the spectral index of the tensor perturbation and that of the scalar perturbation are defined by
\begin{eqnarray}
 r=\frac{\mathcal{P}_T}{\mathcal{P}_\mathcal{R}} \ , 
 \qquad  n_T=\frac{d \log \mathcal{P}_T}{d \log k} \ , 
 \qquad  n_s-1=\frac{d \log \mathcal{P}_\mathcal{R}}{d \log k} \ .
\end{eqnarray}
We can obtain the expressions of them in terms of the slow roll parameters in the usual manner.
\begin{eqnarray}
 r=16\bar{\epsilon} \ , 
 \qquad  n_T=-2\bar{\epsilon} \ , 
 \qquad  n_s-1=-6\bar{\epsilon}+2\bar{\eta} \ .
\label{rntns_mod}
\end{eqnarray}
The consistency relation 
\begin{eqnarray}
 r=-8n_T \ ,
\end{eqnarray}
is not modified from the general relativity case. 
It is straightforward
to relate the slow roll parameters appearing above to those in the
original system which are defined in Eq.~\eqref{slow_param}.

The first derivative of the new potential is calculated as
\begin{eqnarray}
 \bar{V}_\psi=\frac{d}{d\psi}\Omega^4[V+U]=\Omega^3 V_\phi \ ,
\end{eqnarray}
where we have used \eqref{xi_det}, \eqref{phi_psi_rel} and a background equation in \eqref{slow_back}.
Then, the relation between $\bar{\epsilon}$ and $\epsilon$ is
\begin{eqnarray}
 \bar{\epsilon}
=\frac{M_g^2 V_\phi^2}{2\Omega^2(V+U)^2}=\frac{1}{(1+U/V)}\epsilon \ .
\label{corre_eps}
\end{eqnarray}
Similarly, the second derivative of $\bar{V}$ is 
\begin{eqnarray}
 \bar{V}_{\psi\psi}=\frac{d}{d\psi}\bar{V}_{\psi}=-\frac{3\Omega^2 U' V_{\phi}^2}{4 F'(V+U)}+\Omega^2 V_{\phi\phi} \ ,
\end{eqnarray}
followed by the relation between $\bar{\eta}$ and $\eta$,
\begin{eqnarray}
 \bar{\eta}=-\frac{3M_g^2U' V_{\phi}^2}{4\Omega^2 F'(V+U)^2}+\frac{M_g^2 V_{\phi\phi}}{\Omega^2 (V+U)}
=-\frac{3}{2}\frac{U'}{F'}\frac{1}{1+U/V}\epsilon+\eta\ .
%=\frac{1}{(1+U/V)}\biggl[-\frac{3U'}{2F'}\epsilon+\Bigl(1+\frac{U}{V}\Bigr)\eta\biggr] \ .
\label{corre_eta}
\end{eqnarray}
The tensor spectrum and the scalar spectrum~\eqref{power_spectra} are written in terms of the original potential as 
\begin{eqnarray}
 \mathcal{P}_T=\frac{2V}{3\pi^2 M_{\rm eff}^4}\frac{1}{(1+U/V)}
\ , \qquad 
 \mathcal{P}_\mathcal{R}=\frac{V}{24\epsilon \pi^2 M_{\rm eff}^4} \ .
\end{eqnarray}
%\mathcal{P}_\mathcal{R}=\frac{V}{24\pi^2 M_{\rm eff}^4}\frac{1}{(1+U/V)^2} \ .
The resulting tensor to scalar ratio and spectral index are 
\begin{equation}
 r=16 \epsilon \frac{1}{(1+U/V)} \ , \qquad 
 n_T=-2\epsilon\frac{1}{(1+U/V)} \ , \qquad
 n_s-1 =-6\epsilon\Bigl(\frac{1+U'/2F'}{1+U/V}\Bigr)+2\eta \ .
\label{TS_ratio}
\end{equation}
Furthermore, the effective ratio of the slow roll parameters is %modified as
\begin{eqnarray}
 \frac{\bar{\eta}}{\bar{\epsilon}}=-\frac{3U'}{2F'}+\Bigl(1+\frac{U}{V}\Bigr)\frac{\eta}{\epsilon} \ . 
\label{etaeps_mod}
\end{eqnarray}
We evaluate the relation between the tensor to scalar ratio and the spectral index of the scalar spectrum.
%The relation looks like that we obtain in general relativity omitting the bars. 
It is obtained from Eqs.~\eqref{rntns_mod} and \eqref{etaeps_mod} as
\begin{eqnarray}
 n_s-1&=&-\frac{3}{8}r\Bigl(1-\frac{1}{3}\frac{\bar{\eta}}{\bar{\epsilon}}\Bigr)\nonumber\\
 &=&-\frac{3}{8}r\Bigl(1-\frac{1}{3}\frac{\eta}{\epsilon}+\frac{1}{2}\frac{U'}{F'}-\frac{1}{3}\frac{\eta}{\epsilon}\frac{U}{V}\Bigr) \ . %\nonumber\\
% &=&-\frac{3}{8}r\Bigl(1-\frac{1}{3}\frac{\eta}{\epsilon}\biggl[1-\frac{3}{2}\frac{\epsilon}{\eta}\frac{U'}{F'}+\frac{U}{V}\biggr]\Bigr) \ .
\end{eqnarray}
%&=&-\frac{3}{8}r\Bigl(1-\frac{1}{3}\biggl[-\frac{3U'}{2F'}+\Bigl(1+\frac{U}{V}\Bigr)\frac{\eta}{\epsilon}\biggr]\Bigr)\nonumber\\
We find that the slope on the $n_s$-$r$ plane drawn when we vary the 
e-folding number becomes either steeper or
more gradual depending on the sign of
\begin{eqnarray}
 \frac{1}{2}\frac{U'}{F'}-\frac{1}{3}\frac{\eta}{\epsilon}\frac{U}{V} \ .%-\frac{3}{2}\frac{\epsilon}{\eta}\frac{U'}{F'}+\frac{U}{V} \ . 
\label{gradient}
\end{eqnarray}
%than in general relativity.
%when we specify the inflaton potential and $\eta/\epsilon$ is determined. 

\section{Power law potential case}\label{Power-Law}

Here we adopt the power law form $\phi^n$ as the scalar field potential to show
a concrete example. This simple potential form is attractive to lead to a chaotic inflation 
scenario\cite{Linde:1983gd}. However, in the standard inflation based on general
relativity, the observational constraint now 
almost excludes $n=4$ case, and $n=2$ quadratic potential is also in 
tension. The question is whether or not such tension can be relaxed by 
considering the bimetric modification of gravity. 
 
We will find that the tensor to scalar ratio is always larger
than that in general relativity when the e-folding number is fixed though
the spectral index can smaller or larger depending on the details of the
bimetric potential. Therefore, the bimetric modification of gravity 
does not cure the tension mentioned above. 

For $V(\phi)\propto \phi^n$, the ratio of $\eta$ to $\epsilon$ is a constant 
determined only by the power law index $n$:
\begin{eqnarray}
 \frac{\eta}{\epsilon}=\frac{2V V_{\phi\phi}}{V_\phi^2}=\frac{2(n-1)}{n} \ ,
\end{eqnarray}
while the respective slow roll parameters are given by 
\begin{eqnarray}
 \epsilon=\frac{M_{\rm eff}^2}{2}\frac{n^2}{\phi^2} \ , \qquad 
 \eta=M_{\rm eff}^2\frac{n(n-1)}{\phi^2} \ . 
\label{epseta_N}
\end{eqnarray}
The e-folding number, which measures how long inflation continues, is evaluated as 
\begin{eqnarray}
 N= \int_{t}^{t_{\rm e}} H dt =\int^\phi_{\phi_{\rm e}} \frac{3H^2}{V_\phi}d\phi=\int^\phi_{\phi_{\rm e}}\frac{V}{V_\phi M_{\rm eff}^2}d\phi \ ,
\end{eqnarray}
where $t_{\rm e}$ and $\phi_{\rm e}$ are the time and the field value at the end of inflation.
The dependence of $\xi$ on $\phi$ is found in \eqref{logO_phi}, which means that $\xi$ moves slower than $\phi$ as $\mathcal{O}(\epsilon^{1/2})$ since we assumed that $V/M_g^2 W$ is not enhanced.
%\begin{eqnarray}
% \frac{d\xi}{d\phi}=\frac{V_\phi}{F'}=-\frac{V_\phi \xi}{3M_g^2W}=-\sqrt{2\epsilon}\xi\frac{ M_{\rm eff}}{M_g^2}\frac{H^2}{W} \ ,
%\end{eqnarray}
%where we used the background equations, the definition of $\epsilon$ and the relation (\ref{rel_FW}). 
%As we mentioned in Sec.~\ref{Cosmo_Pert}, we have assumed that the Higuchi bound is satisfied, $W>0$, $d\xi/d\phi<0$. 
%$d\xi/d\phi$ is slow roll suppressed and 
%Unless the Higuchi bound is saturated $W\ll H^2$, 
Therefore, the effective gravitational constant can be approximated 
by a constant in the integral over $\phi$. 
Then, the e-folding number is given by
\begin{eqnarray}
 N= \frac{\phi^2}{2nM_{\rm eff}^2} \ .
\end{eqnarray}
If we eliminate $\phi^2$ from the expressions of $\epsilon$ and $\eta$~\eqref{epseta_N} using this relation, we gain
\begin{eqnarray}
 \epsilon=\frac{n}{4N} \ , \qquad 
 \eta=\frac{n-1}{2N} \ ,
\end{eqnarray}
which are the same as those in general relativity. %From Eqs. (\ref{rntns_mod}) and (\ref{corre_eps}), the tensor to scalar ratio is written as
From \eqref{TS_ratio}, we find
\begin{eqnarray}
 r=\frac{4n}{N}\frac{1}{(1+U/V)}>r_{GR} \ ,
\end{eqnarray}
where $r_{GR}=4n/N$. 
The inequality is verified under 
a reasonable assumption that Minkowski spacetime is
realized in the low-energy regime, which means $H^2=0$ when the matter energy density is zero, $\rho=0$. 
This implies $U=U'=0$ at $\rho=0$ from Eq.~\eqref{Friedmann}. We also 
find that $U'$ is positive from Eq.~\eqref{Friedmann} as long as
$H^2>0$. 
Combining this with~\eqref{Higuchi_condition}, equivalent to $d\rho/d\xi<0$, we find that 
\begin{equation}
 U<0 \ ,
\label{UVsignature}
\end{equation} 
for any finite positive $\rho$ in the healthy branch.
The positivity of $U+V$, which is also verified from 
Eq.~\eqref{Friedmann}, guarantees the positivity of $1/(1+U/V)$ and it is
larger than unity. Consequently, the tensor to scalar ratio in bimetric
theory is larger than $r_{GR}$ for a fixed e-folding number. 
For the same reason, we also have a larger absolute value for the spectral index of tensor perturbation compared with the case of general relativity as you can see in Eq.~\eqref{TS_ratio}.
It is surprising that such an inequality holds irrespective of 
the choice of the model parameters~$\{c_k\}$. 

According to \eqref{TS_ratio}, the difference of the spectral index from the general relativity case is
also evaluated as
\begin{eqnarray}
 n_s-1
 &=&(-6\epsilon+2\eta)+6\epsilon\,\frac{1}{1+U/V}\Bigl(-\frac{1}{2}\frac{U'}{F'}+\frac{U}{V}\Bigr) \nonumber \\
 &=&(n_s-1)_{GR}+\frac{3n}{2N}\frac{1}{1+U/V}\Bigl(-\frac{1}{2}\frac{U'}{F'}+\frac{U}{V}\Bigr) \ ,
\end{eqnarray}
%&=&-\frac{3}{8}\frac{4n}{N}\frac{1}{(1+U/V)}\Bigl(1-\frac{1}{3}\biggl[-\frac{3U'}{2F'}+\Bigl(1+\frac{U}{V}\Bigr)\frac{2(n-1)}{n}\biggr]\Bigr)\nonumber\\
% &=&\frac{n-1}{N}-\frac{3n}{2N}\frac{1}{1+U/V}\Bigl(1+\frac{1}{2}\frac{U'}{F'}\Bigr)\nonumber\\
where $(n_s-1)_{GR}=-(n+2)/2N$. It depends on the sign of 
\begin{eqnarray}
  -\frac{1}{2}\frac{U'}{F'}+\frac{U}{V} \ ,
\end{eqnarray}
whether the spectral index is larger or smaller than the general relativity counterpart.

The slope on the $n_s$-$r$ plane when the e-folding number is varied 
is steeper or more gradual compared with the general relativity case
depending on the sign of the expression~\eqref{gradient}, 
which in the present case reduces to 
\begin{eqnarray}
 \frac{1}{2}\frac{U'}{F'}-\frac{1}{3}\frac{\eta}{\epsilon}\frac{U}{V}
 =\frac{1}{2}\frac{U'}{F'}-\frac{2(n-1)}{3n}\frac{U}{V} \ .
 %-\frac{3}{2}\frac{\epsilon}{\eta}\frac{U'}{F'}+\frac{U}{V}=-\frac{3n}{4(n-1)}\frac{U'}{F'}+\frac{U}{V} \ .
\label{slope}
\end{eqnarray}
%corresponds to the case that the slop becomes gradual. 
When this expression is positive, the slope becomes more gradual. At that time, 
the modification to $n_s-1$ is evaluated as 
\begin{eqnarray}
 -\frac{1}{2}\frac{U'}{F'}+\frac{U}{V}<\frac{(n+2)}{3n}\frac{U}{V}<0 \ ,
\end{eqnarray}
since $U<0$ and $V>0$.
Therefore, $n_s-1$ is smaller than the general relativity case. 

On the other hand, the slope becomes steeper when 
the expression~\eqref{slope} is negative. 
In this case, we can give a lower bound on the modification of $n_s$ as 
\begin{eqnarray}
(n_s-1)-(n_s-1)_{GR}=
 \frac{3n}{2N}\frac{1}{1+U/V}
 \left(
-\frac{1}{2}\frac{U'}{F'}+\frac{U}{V}\right)
  >
 \frac{3n}{2N}\frac{1}{1+U/V}
 \left(
\frac{(n+2)}{3n}\frac{U}{V}\right) \ .\nonumber\\
\end{eqnarray}
However, the right hand side of the above inequality 
is negative. Therefore, the sign is indefinite and $n_s-1$ can either
decrease or increase. % (See Fig. \ref{})

%In the case of $V\propto \phi$, the expression simplifies a little since
%$\eta$ vanishes. In this case, the relation between $n_s$ and $r$ is
%\begin{eqnarray}
%  n_s-1=-\frac{3}{8}r\Bigl(1+\frac{1}{2}\frac{U'}{F'}\Bigr)  \ .
%\end{eqnarray}
%Since %$n_s-1=-3r/8$ and 
%\begin{eqnarray}
% \frac{1}{2}\frac{U'}{F'}<0, 
%\end{eqnarray}
%the slope becomes steeper, but $n_s-1$ can either decrease or increase.

%\begin{figure}[!h]
%\centering \includegraphics[height=3.5in]{slide.jpg}
%\caption{}
%\label{fig. ch5}
%\end{figure}

\section{Summary}

Introducing an approximation, which is based on the decay of the massive gravitons and is valid at the leading order in slow roll, we calculated the spectrum of the tensor perturbation and that of the scalar perturbation generated during inflation. Under the approximation, the action up to quadratic order in perturbations reduces to the Einstein theory with a non-minimally coupled inflaton field which has a modified potential. After conformal transformation, the spectra are easily obtained in the standard manner. We found how the tensor to scalar ratio and the spectral index of the scalar perturbation are modified for general choice of model parameters in dRGT bimetric gravity except when the Higuchi bound is saturated.
As a concrete example, we examined the power law potential cases.
%and evaluated the tensor to scalar ratio and the spectral index. 
Opposed to the naive expectation, the tensor to scalar ratio is larger than the general relativity counterpart in general, while the spectral index can be either larger or smaller than that in general relativity, depending on e-folding number, power law index of the inflaton potential and the parameters in the bimetric interaction. The tensor to scalar ratio is observationally constrained from above, which is already in tension if we adopt simple power law potentials of inflaton~\cite{Ade:2015lrj,Ade:2015tva,Array:2015xqh}. Therefore, it becomes more difficult to be consistent with observations when we consider bimetric extension of gravity. Nevertheless, this fact does not immediately exclude dRGT bimetric theory since the correction can be small enough depending on model parameters of bimetric theory. 
We also saw that the consistency relation is not modified and the spectral index of the tensor perturbation has a negative larger value in this theory than in general relativity.
%The reason why the tensor amplitude is enhanced relative to the scalar amplitude is that the effective gravitational constant which determines the normalization of the gravitational perturbations,
%is different from that determined by the background evolution,

We would like to mention again that we have assumed the coupling constant, $m$, to be larger than the expansion rate after inflation, $H$, to avoid gradient instability. The case $m\gg H$ corresponds to the low energy limit, in which, for instance, $U/V$ converges to $0$ as $\xi-\xi_0$. The effective mass is proportional to the coupling constant $m$ and has the same order with it unless the parameters are tuned. Then, the bimetric effects in the low energy regime are suppressed due to the large graviton mass. We can also keep the effective mass small in the low energy regime by tuning the parameters~$\{c_k\}$, and then observable effects due to the presence of massive gravitons can appear.

The reason why the tensor amplitude is enhanced relative to the scalar amplitude can be more intuitively understood as follows.
The effective gravitational coupling for the pure gravity sector,
\begin{equation}
 G^{(T)}=\frac{1}{M_g^2(1+\kappa\xi^2)} \ ,
\end{equation}
which determines the tensor amplitude, is different from the effective gravitational coupling read off from the Friedmann equation, 
\begin{equation}
 G^{(S)}=\frac{1+U/V}{M_g^2(1+\kappa\xi^2)} \ ,
\end{equation}
which determines the scalar amplitude, in dRGT bimetric theory, while they coincide with each other in general relativity. As discussed in Sec.~\ref{Power-Law}, under our assumption that Minkowski spacetime is realized in the low-energy regime, the bimetric correction to the potential of the inflaton $U$ is always negative and hence $1+U/V$ is smaller than unity. As a result, we have $G^{(T)}>G^{(S)}$, which leads to the amplification of the tensor to scalar ratio. Consistency of the results with late time evolution of the universe and the perturbation should be discussed as a succeeding work.

\acknowledgements
We would like to thank Antonio de Felice, Shinji Mukohyama and Jiro Soda for fruitful discussions. 
YS is supported by the Grant-in-Aid for Scientific Research (No.~24103006).
%the Grant-in-Aid for Japan Society for the Promotion of Science(JSPS) Fellows No. 261236. 
TT is supported by the Grant-in-Aid for Scientific Research (Nos.~24103006,
24103001, 26287044 and 15H02087).

%\appendix

%\bibliographystyle{JHEP}
%\bibliography{references}
\bibliography{BiTN_JCAP}

\end{document}